\documentclass[aps,prb, reprint, superscriptaddress]{revtex4-1}

\usepackage{graphicx}
\usepackage[hidelinks]{hyperref}

\usepackage{color}
\usepackage{soul}
\usepackage{dcolumn}
\usepackage{amsmath}
\usepackage{amssymb}
\usepackage{bm}

\newcommand{\vlad}[1]{\textcolor{magenta}{#1}}

\begin{document}

\title{\vlad{Blochnium: the superconducting quasicharge qubit}}

\title{The superconducting quasicharge qubit}

\author{Ivan V. Pechenezhskiy}
\thanks{These authors contributed equally.}
\author{Raymond A. Mencia}
\thanks{These authors contributed equally.}
\author{Long~B.~Nguyen}
\author{Yen-Hsiang Lin}
\author{Vladimir E. Manucharyan}
\affiliation{Department of Physics, Joint Quantum Institute, and Quantum Materials Center, University of Maryland, College Park, MD 20742, USA}

\date{\today}

\maketitle

\textbf{The non-dissipative non-linearity of a~Josephson junction~\cite{josephson1962possible} converts macroscopic superconducting circuits into artificial atoms~\cite{clarke1988quantum}, 
enabling some of the best controlled quantum bits (qubits) today~\cite{devoret2013superconducting, Google}. 
Three fundamental types of superconducting
qubits are known~\cite{clarke2008superconducting}, each reflecting a distinct behavior of quantum fluctuations in a Cooper pair condensate: single charge tunneling (charge qubit~\cite{nakamura1999coherent, vion2002manipulating}), single flux tunneling (flux qubit~\cite{chiorescu2003coherent}), and phase oscillations (phase qubit~\cite{martinis2002rabi}/transmon~\cite{koch2007charge, schreier2008suppressing}). Yet, the dual nature of charge and flux suggests that circuit atoms must come in pairs.
Here we introduce the missing one, named ``blochnium". It exploits a coherent insulating 
response
of a single Josephson junction that emerges from the extension of phase fluctuations 
beyond the $2\pi$-interval~\cite{schmid1983diffusion, bulgadaev1984phase, averin1985bloch, schon1990quantum}. 
Evidence for such effect was found in an~out-of-equilibrium dc-transport through junctions connected to high-impedance
leads~\cite{kuzmin1991observation, Haviland96, penttila1999superconductor, watanabe2001coulomb, corlevi2006phase},
although 
a~full consensus is absent to date~\cite{ergul2013localizing, cedergren2017insulating, murani2019absence}.
We shunt a~weak junction with an exceptionally high-value inductance --- the key technological innovation behind our experiment --- and measure the rf-excitation spectrum as a function of external magnetic flux through the resulting loop. 
The junction's insulating character manifests
by the vanishing flux-sensitivity of the qubit transition between the ground and the first excited states, which nevertheless rapidly recovers for transitions to higher energy states. 
The spectrum agrees with a duality mapping of blochnium onto transmon, which replaces the external flux by the offset charge and introduces a new collective quasicharge variable in place of the superconducting phase~\cite{matveev2002persistent, koch2009charging}. Our result unlocks the door to
an unexplored regime of macroscopic 
quantum dynamics in ultrahigh-impedance 
circuits, which may have applications to quantum computing and quantum metrology of direct current.}


Is a Josephson tunnel junction between two superconductors a superconducting link or an insulating break? Josephson showed that a~junction can be viewed as a~non-linear inductance that carries flux $\hbar/2e \times \varphi$ and energy $E = -E_J\cos\varphi$, where $\varphi$ is the superconducting phase-difference, $\hbar$ is the reduced Plank's constant, $2e$ is the Cooper pair charge, and $E_J$ is the Josephson energy. If quantum fluctuations of $\varphi$ are small compared to $2\pi$,
the inductance can be linearized and the junction responds as a superconductor (Fig.~1a). Yet, an~opposite scenario was suggested in the case $\varphi$ is free to extend beyond the $2\pi$-interval~\cite{averin1985bloch}. In what follows, it is essential to take into account the junction's intrinsic oxide capacitance across the Josephson element (Fig.~1b).
The resulting circuit equations mimic an electron in a crystal (Fig.~1b): the flux 
is the position, the capacitance is the mass, the charge on the capacitor is the momentum, while the Josephson energy corresponds to a periodic crystal field (Fig.~1c).
The dynamics of $\varphi$ can be described by extended Bloch waves and continuous Bloch bands. The energy $E_B(q)$ within the lowest band would be a $2e$-periodic function of the circuit analog of quasimomentum --- the \textit{quasicharge}~$q$. Connecting a Cooper pair tunneling to a Bragg reflection, the quasicharge is the externally 
supplied charge. In other words, at low frequencies the junction transforms into a non-linear Bloch capacitance (an equivalent of the effective mass) that stores charge $q$ and is characterized by a~$2e$-periodic charging energy $E_B(q)$. When quantum fluctuations of $q$ are suppressed, the Bloch capacitance can be linearized, and hence the junction responds as an insulator.

\begin{figure*}
	\centering
	\includegraphics[width=0.98\linewidth]{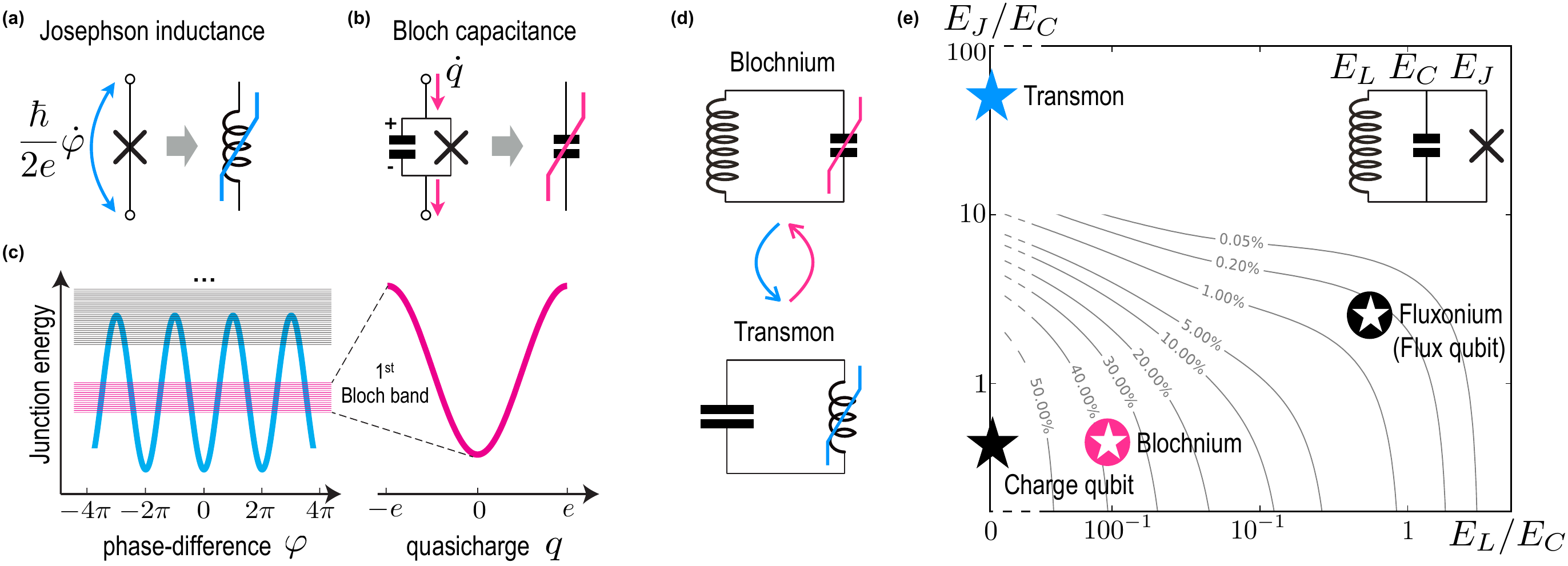}
	\caption{\textbf{Blochnium artificial atom.} (a) A~Josephson junction is a non-linear inductance storing flux $(\hbar/2e)\varphi$. (b)~A~junction shunted by a small linear capacitance becomes a non-linear Bloch capacitance storing quasicharge $q$. (c) Spectrum of Bloch bands (magenta, gray) originating from the quantum motion of $\varphi$ in the periodic Josephson potential (blue). The first Bloch band energy is a $2e$-periodic function of quasicharge $q$ (magenta) and it defines the charging energy of the Bloch capacitance. (d)~Blochnium circuit concept (top) and its dual transmon circuit concept (bottom). (e) Parameter space of the four fundamental qubits, all defined by the same three-element circuit (inset) with vastly different combinations of $E_L$, $E_C$, and $E_J$ (see text). The contours show the calculated probability of $|\varphi|>\pi$.
}
\end{figure*}

The junction's external circuit plays a decisive role in choosing between the two antagonistic scenarios.
The Bloch oscillations at the frequency $I/2e$ are expected in response to a dc-current $I = \dot{q}$ driven through the Bloch capacitance by an infinite-impedance current source~\cite{averin1985bloch}.
By contrast, the Josephson oscillations at a frequency $2eV/\hbar$ are induced in response to a dc-voltage $V = (\hbar/2e)\dot{\varphi}$ across the junction, applied with a zero-impedance voltage source. 
We short-circuit a Bloch capacitance with a large-value linear inductance.
The resulting non-linear and non-dissipative electrical circuit is ``blochnium" artificial atom (Fig.~1d, top). Blochnium is the dual of transmon, a Josephson inductance shunted by a large-value linear capacitance  (Fig.~1d, bottom). The high- (low-) impedance linear circuit element in blochnium (transmon) suppresses quantum fluctuations of the $q-$ $(\varphi-)$ variable and thereby stabilizes the insulating (superconducting) 
behavior of the junction.
Quasicharge localization profoundly differs from the usual Coulomb blockade: quasicharge cannot be offset from the mean value $q=0$ by a static electric field thanks to complete screening by the galvanic shunt. 
The low-energy excitations of blochnium are anharmonic vibrations of quasicharge through the small junction, 
the spectrum of which we measured for the first time.

Constructing blochnium amounts to choosing three elementary energy scales: the Josephson energy $E_J$, the charging energy $E_C = e^2/2C$ of the total capacitance $C$ across the Josephson element, and the inductive energy $E_L = (\hbar/2e)^2/L$ associated with storing a flux quantum in the inductance $L$~(Fig.~1e). Our devices have $E_J/E_C \approx 1$ and $E_L/E_C \approx 1/100$. The first condition conveniently maximizes the width of the lowest Bloch band along with the gap to the next one. As for the second condition, in the special case $E_L=0$, we are left with two isolated grains linked by a single Cooper pair tunneling, i.e. a charge qubit. In such a case, the quasicharge looses dynamics and it can be interpreted as an external charge-offset. For $E_L/E_C \lesssim 1$, we get fluxonium~\cite{Manucharyan113, nguyen2018high}, a high-coherence implementation of a~flux qubit, where a superconducting loop is disrupted by the tunneling of a single flux quantum. 
In a~charge qubit, reducing $E_C$ proliferates multiple Cooper pair tunneling\cite{koch2007charge}, which establishes a well-defined phase-difference across the junction due to the Heisenberg uncertainty principle.
This is how a charge qubit evolves into transmon for $E_J/E_C \sim 10-100$. Blochnium emerges from the dual evolution of a flux qubit/fluxonium on reducing $E_L$.
The probability to find $\varphi$ outside a single Josephson well becomes significant for $E_L/E_C \lesssim 1/100$ (Fig.~1e).

\begin{figure*}
	\centering
	\includegraphics[width=0.98\linewidth]{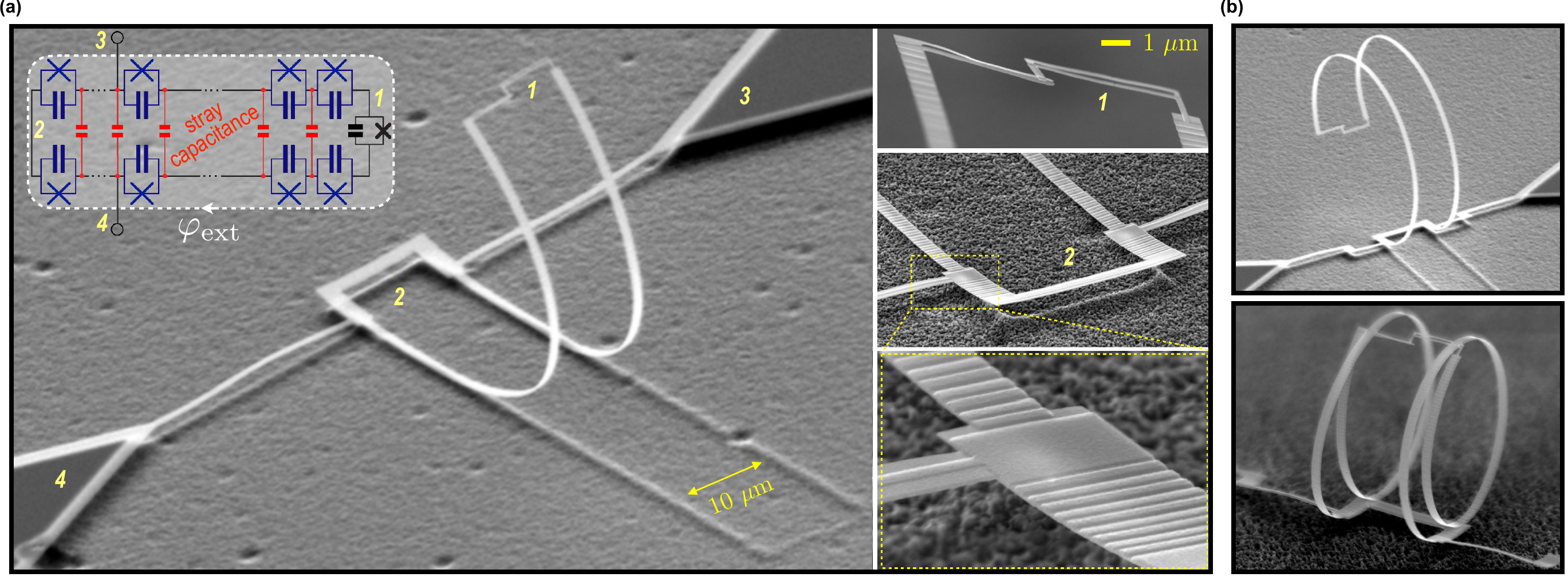}
	\caption{\textbf{Blochnium implementation.}
		(a) Scanning electron micrographs of a fabricated device released from the substrate in order to reduce the stray capacitance. The released Josephson chain curls upward and elevates the small junction by a~few tens of micrometers above the substrate. The inset shows a circuit model of the device in the form of a superconducting loop interrupted by a small-area junction (black) and the larger-area chain junctions (dark blue). The stray capacitances are marked in red. The indexes $1 - 4$ mark the small junction, the opposite end of the loop, and the two connections to the readout circuitry [not shown in the image]. External magnetic field induces flux $\varphi_{\textrm{ext}}$ through the loop. (b, c) Examples of released blochnium circuits with different degrees of curling. The two parts of the Josephson junction chain are spaced by 10~$\mu\mathrm{m}$ in all devices.
	}
\end{figure*}

One of our measured devices is shown in Fig.~2a. Building on fluxonium results, we constructed a compact shunt using the kinetic inductance of a
Josephson junction chain (Fig.~2a-inset, blue). The key innovation here is that we released the entire circuit from the substrate and suspended it in vacuum (Fig.~2a). 
With optimally chosen junction parameters, Josephson chains can reach an exceptionally high inductance density of $10^4\mu_0$ ($\sim10~\textrm{H}/\textrm{m}$), where $\mu_0$ is vacuum permeability, before the detrimental effects associated with the superconductor-insulator transition kick in~\cite{manucharyan2012evidence, kuzmin2018quantum}. However, the total inductance 
is also limited by the chain self-capacitance, originating from the stray electrostatic coupling between the opposite-facing metal islands (Fig.~2a inset, red). Besides introducing parasitic modes, the self-capacitance contributes to $C$ and this effect prevents reducing $E_L/E_C$ far below unity by lengthening the chain. The stray capacitance, however, is unnecessarily large in most superconducting circuits due to the high relative dielectric permittivity of silicon ($\epsilon\approx12$) or sapphire ($\epsilon \approx10$), the common low-loss substrate materials. By eliminating the substrate contribution, we reduced the stray capacitance nearly tenfold, which provided the required leap in the inductance value.

The substrate-free blochnium devices in Fig.~2 are created in a~two-step fabrication process. First, a superconducting loop with up to $460$ Al/AlOx/Al chain junctions and one small junction is fabricated using the standard Dolan bridge technique. Next, a gentle burst of isotropic silicon etch is applied, with the oxidized Al film acting as a natural mask. Because silicon etching is more efficient underneath the skinnier leads, the small junction end of the chain (labeled~`1' in Fig.~2a) detaches from the substrate prior to the other parts and immediately curls upwards driven by the strain relaxation. The curling effect is robust and reproducible. Moreover, it is possible to vary the amount of curling (Fig.~2b,c). We focused on devices with a nearly vertically standing chain (Fig.~2a), where parasitic capacitance is minimal.

The loop is inductively coupled to the readout circuitry following a previously developed method~\cite{kou2018simultaneous}. A small section of the loop (labeled `2' in Fig.~2a) is connected to a capacitive antenna (whose leads are labeled `3' and `4' in Fig.~2a) which forms a readout resonator for coupling the device to the measurement apparatus. The transition spectrum as a function of the flux through the loop  (Fig.~3a) was measured using a conventional two-tone rf-spectroscopy~\cite{schuster2005ac} (see Supplementary Information for spectroscopy details). To identify transitions, we compare the data (Fig.~3a, markers) to the spectrum of a three-element circuit (Fig.~1e) Hamiltonian
\begin{equation}
H_1 = 4 E_C (Q/2e)^2 + \frac12 E_L \varphi ^2 - E_J \cos (\varphi-\varphi_{\textrm{ext}}).
\end{equation}
The Hamiltonian~(1) describes a particle that is both in a flux-tuned periodic potential and a soft harmonic trap due to the $E_L$-term. The operators $\varphi$ and $Q$ obey the position-momentum type commutation relation $[\varphi, Q/2e] = i$.  The simple model of Hamiltonian~(1) accurately fits the lowest five transitions (Fig.~3, dashed lines). The fit parameters $E_J/h = 4.70\;\textrm{GHz}$, $E_C/h = 7.07\;\textrm{GHz}$, and $E_L/h = 66.5\;\textrm{MHz}$ indeed define a previously inaccessible spot on the circuit parameter map of Fig.~1e ($E_J/E_C =0.66$,  $E_L/E_C = 0.009$). The capacitance $C \approx 2.7~\textrm{fF}$ can be almost entirely accounted for by the small junction oxide capacitance. The inductance $L \approx 2.5~\mu \textrm{H}$ exceeds that of a typical fluxonium tenfold while showing no influence of 
parasitic modes within the entire frequency range of Fig.~3.

\begin{figure}
	\centering
	\includegraphics[width=0.98\linewidth]{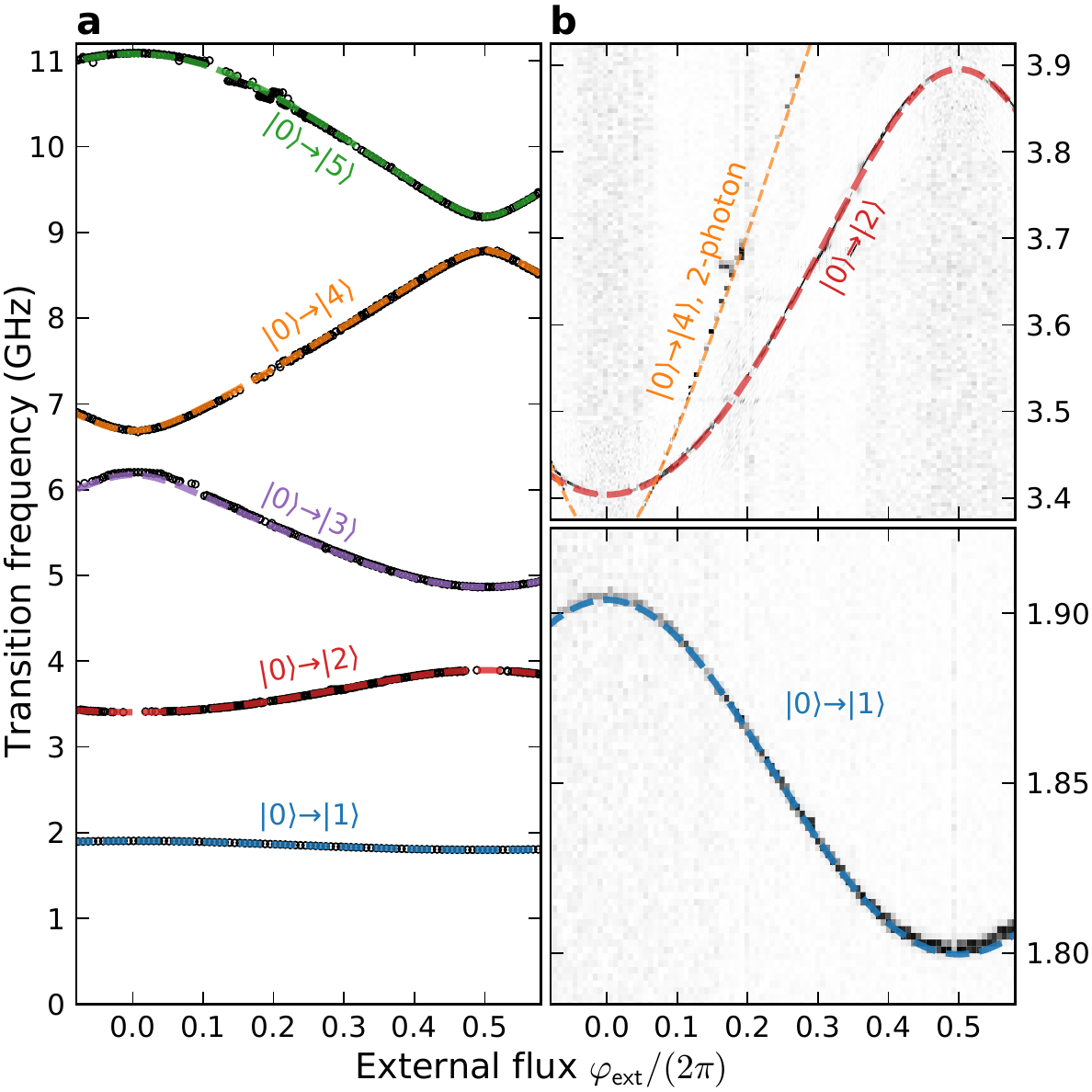}
	\caption{\textbf{Measured transitions of blochnium.} (a)~Transition frequencies (black markers) extracted from the two-tone spectroscopy data as a~function of the external flux through the loop and the fit (dashed lines) to the spectrum of Hamiltonian~(1). (b) Raw data zoom-in on the $|0\rangle\rightarrow|2\rangle$ and the two-photon $|0\rangle\rightarrow|4\rangle$ transitions (top) and the qubit transition $|0\rangle\rightarrow|1\rangle$ (bottom). Note the qubit flux-modulation is only about $100~\textrm{MHz}$.
	}
\end{figure}

The rapid crossover in the transition's flux-modulation characteristic from a weak harmonic to a strong saw-tooth (Fig.~3)
has no analogs among previously reported spectra of superconducting quantum interference devices. 
Let us introduce a phenomenological model of an inductively-shunted Bloch capacitance (Fig.~1d), whose Hamiltonian is
\begin{equation}
H_2 = 2\pi^2E_L(m - \varphi_{\textrm{ext}}/2\pi)^2 + E_B(q).
\end{equation}
Here the quasicharge $q$ transferred across the shunt is a compact variable at the interval $(-e,e]$ and $E_B(q)$ is the $2e$-periodic charging energy of the Bloch capacitance. The conjugate momentum $m$ is an integer operator satisfying $\exp(- i \pi q/e) m \exp( i\pi q/e)  = m +1$. The external flux $\varphi_{\textrm{ext}}$ couples to momentum like a gauge field. For a sufficiently large $E_L$ the momentum $m$ counts the flux quanta (or the $2\pi$-slips of phase) in the loop, although such a notion becomes progressively more vague upon reducing $E_L$ into the regime of this work's interest.

Aside from the unessential effect of higher harmonics of $E_B(q)$, Hamiltonian~(2) models a quantum pendulum with the deflection angle given by $\pi q/e$. The same pendulum model describes transmons, where the deflection angle is $\varphi$ and the external flux is replaced by the offset charge, thereby providing a quantitative basis for the duality. Using the circuit parameters $E_J$ and $E_C$, extracted above by fitting the Hamiltonian~(1) model to the data, we calculate dispersion of the Bloch bands originating from the small junction in our circuit. The function $E_B(q)$ in Hamiltonian~(2) is set to be the lowest band, and the higher bands are disregarded. Using the Hamiltonian~(1) fit value for $E_L$, we numerically diagonalize the Hamiltonian (2) and use both models to interpret the spectroscopy results (Fig.~4).




The lowest five energy states exhibit all the essential features of a transmon device~\cite{koch2007charge} if one pretends that $\varphi_{\textrm{ext}}/2\pi$ is the offset charge normalized by $2e$ (Fig.~4a). The ground state $|0\rangle$ has a vanishing flux-dispersion (unresolvable in Fig.~4a), corresponding to a hardly measurable persistent current $7~\textrm{pA}$ (Supplementary Information). The $|0\rangle$ energy level lies deeply inside the first Bloch band (Fig.~4b), and this property links the absence of magnetic field response to the localization of quasicharge around $q=0$.
Indeed, with $\langle 0|(q/2e)^2|0\rangle \approx 0.019$, the quasicharge wavefunction has exponentially small 
tails at the Brillouin zone edges $|q|=e$  (Fig.~4c), in which case $\varphi_{\textrm{ext}}$ can be eliminated from Hamiltonian (2) by a~gauge transformation $\exp(i\varphi_{\textrm{ext}}q/2e)$. The $|1\rangle$ level also lies well inside the first Bloch band, having only $5\%$ of flux modulation due to a reduced quasicharge confinement. The $|0\rangle \rightarrow|1\rangle$ transition corresponds to semi-classical oscillations of quasicharge inside the Bloch band potential.
The flux quantization recovers already for states $|3\rangle$ and $|4\rangle$ as the quasicharge spills over the entire Brillouin zone. At higher energies, the spectra of Hamiltonians~(1) and (2) deviate due to the presence of higher Bloch bands, ignored in the Hamiltonian~(2) model  (Fig.~4b). In fact, the quantitative discrepancy testifies that our Bloch capacitance emerges from the physics of the underlying Josephson effect.


\begin{figure}
	\centering
	\includegraphics[width=\linewidth]{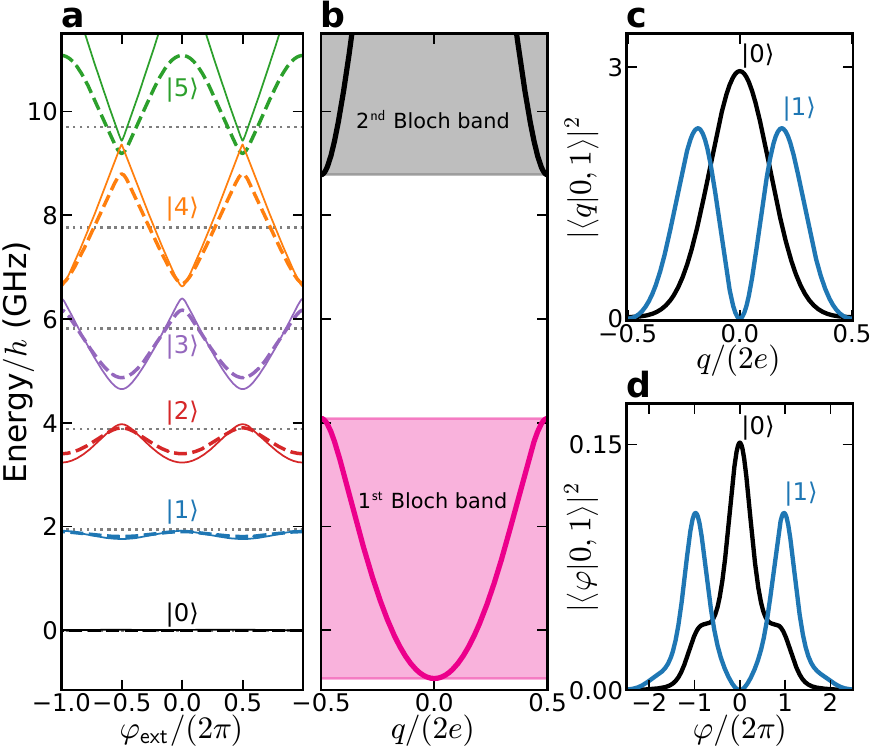}
	\caption{\textbf{
	Interpretation of blochnium spectroscopy.
	} (a) Eigenenergies of Hamiltonian~(1) (dashed lines) and Hamiltonian~(2) (solid lines), calculated using the extracted device parameters vs. the external flux $\varphi_{\textrm{ext}}$. Dotted lines indicate the spectrum of Hamiltonian~(1) for $E_J=0$. (b)~Calculated energies in the lowest two Bloch bands as a function of the quasicharge $q$. (c,d)~Ground and first excited states probability distributions at $\varphi_{\textrm{ext}}=0$ in the (c) quasicharge representation and in the (d) phase-difference representation.
	}
\end{figure}

Quasicharge localization with the r.m.s. value of 13\% of a~Cooper pair (Fig.~4c) justifies the Bloch band picture, in which the junction responds as a Bloch capacitance rather than a Josephson inductance. 
The phase-difference $\varphi$ across the junction extends beyond a single Josephson well but it remains localized at the scale of a few wells (Fig.~4d). Thus, the $\varphi$-variable is no longer compact at $(-\pi, \pi]$, and its localization length continually increases with $L$. Notably, the quantity $\langle 0|\cos\varphi|0\rangle$ remains non-zero in the limit $L \rightarrow \infty$. This means that, while the junction responds as a capacitance, Cooper pairs can virtually tunnel back and forth across the oxide. 
Such processes are, in fact, responsible for the non-linearity of the Bloch capacitance $C_B (q) = (d^2 E_B(q)/dq^2)^{-1}$.
In addition, virtual Cooper pair tunneling would increase $C_B(0)$ significantly above $C$ in junctions with $E_J \gg E_C$. 
In our device, $C_B(0) \approx C$, which allows a straightforward illustration of the junction's insulating character. Namely, setting $E_J = 0$ and keeping the fit values of $E_C$ and $E_L$ reproduces, with a few percent accuracy,
both the $|0\rangle \rightarrow |1\rangle$ transition frequency and the matrix element $\langle 0|\varphi|1\rangle$ (Supplementary Information). In other words, the low-energy dynamics of our device is consistent with simply removing the Josephson "cross" element from the circuit.

Achieving the fluctuations regime of the \textit{decompactified} phase-difference and the localized quasicharge allowed us to complete the table of fundamental superconducting artificial atoms with blochnium. The finite $L$-shunt eliminates the offset-charge sensitivity of blochnium transitions, while the extension of phase fluctuations beyond the $(-\pi, \pi]$-interval renders the $|0\rangle\rightarrow|1\rangle$ qubit transition practically unaffected by the background level of the $1/f$ flux noise. Moreover, transitions to non-computational states are flux-tunable and anharmonic, which is a desirable resource for quantum engineering. Initial time-domain measurements of our substrate-free devices revealed the energy relaxation time $T_1 \approx 10~\mu\textrm{s}$ and the relaxation-limited spin-echo coherence time $T_2 \approx 20~\mu\textrm{s}$ (Supplementary Information).

Blochnium qubit is enabled by a remarkable circuit element to which we refer as ``hyperinductance": a lossless linear inductance $L \approx 2.5~\mu\textrm{H}$ operating beyond the frequency $\omega/2\pi \approx 13\;\textrm{GHz}$, such that $L\omega > 200\;\textrm{k}\Omega$. This impedance value is a factor of 30 greater than the resistance quantum for Cooper pairs $h/(2e)^2$, and it is likely the highest characteristic impedance an electromagnetic structure attained so far. Among other applications, hyperinductance has long been sought after for realizing fault-tolerant logical operations on superconducting qubits~\cite{bell2012quantum, douccot2012physical, brooks2013protected} and for
implementing the quantum current standard via Bloch oscillations~\cite{nguyen2007current, di2015quantum}.

\section*{Materials and Methods}

Devices were fabricated using the conventional Dolan-bridge technique~\cite{dolan1977offset,Frunzio05}.
A bilayer of MMA/PMMA was spun on top of a high-resistivity ($\rho>10\,\textrm{k}\Omega\cdot\textrm{cm}$) silicon substrate covered by the native oxide prior to patterning the device with e-beam lithography. The patterned device was then loaded into a e-beam evaporator where the Al/AlOx/Al Josephson junctions were created by a double angle shadow evaporation of aluminum with an intermediate static oxidation step in between. The first layer of aluminum was 20~nm while the second layer of aluminum was 40~nm. After deposition the resist bilayer was lifted off in a $60^{\circ}\textrm{C}$ acetone bath.
 
Once each device was created in the two-dimensional fashion, we performed an etch to release the aluminum circuit from the silicon substrate. The device was etched using a xenon-difluoride reactive ion etching technique. This etching process relies on the dry etchant selectivity of silicon over aluminum~\cite{Chang95,chu2016suspending}.
Each sample was etched for two minutes at a pressure of 2~mTorr. No noticeable detrimental effects were observed in the scanning electron micrographs of our etched devices after utilizing this fabrication technique.

\section*{Acknowledgments}

R.A.M. fabricated devices and performed measurements guided by I.V.P. I.V.P. analyzed the data and co-wrote the manuscript with V.E.M. L.B.N. and Y.-H.L. built the low-temperature microwave measurement setup. V.E.M. managed the project. All authors contributed to discussions of the results. We acknowledge funding from NSF-DMR (CAREER Award), NSF PFC at JQI, and the ARO-LPS HiPS program.

\vfill
\bibliographystyle{naturemag}
\providecommand{\noopsort}[1]{}\providecommand{\singleletter}[1]{#1}%

\end{document}